\documentstyle[12pt,twoside,fleqn,espcrc1,epsfig]{article}

\newcommand{\bm}[1]{\mbox{\boldmath $#1$}}
\newcommand{\bt}[1]{{\bm #1}_{_T}}
\newcommand{\Pslash}{\kern 0.2 em P\kern -0.56em \raisebox{0.3ex}{/}}
\newcommand{\pslash}{\kern 0.2 em p\kern -0.4em /}
\newcommand{\nslash}{\kern 0.2 em n\kern -0.4em /}
\newcommand{\kslash}{\kern 0.2 em k\kern -0.45em /}
\newcommand{\Sslash}{\kern 0.2 em S\kern -0.56em \raisebox{0.3ex}{/}}
\newcommand{\dslash}{\kern 0.2 em \partial\kern -0.56em \raisebox{0.3ex}{/}}
\newcommand{\xbj}{x_{_B}}

\newcommand{\bkt}{\bt{k}}

\newcommand{\bSt}{\bt{S}}
\def\be{\begin{equation}}
\def\ee{\end{equation}}
\def\bea{\begin{eqnarray}}
\def\ba{\begin{eqnarray}}
\def\eea{\end{eqnarray}}
\def\ea{\end{eqnarray}}
\def\st{{\scriptstyle T}}

\newcommand{\AmS}{{\protect\the\textfont2
A\kern-.1667em\lower.5ex\hbox{M}\kern-.125emS}}

\title{
\begin{flushright}
\small
hep-ph/9907356 
\\
VUTH 99-16
\end{flushright}
\mbox{}\\
Perspectives in Polarized Leptoproduction\footnote{
invited talk at the Workshop on The Structure of the Nucleon
(NUCLEON99), Frascati, June 7-9, 1999}
}

\author{\underline{P.J. Mulders} and M. Boglione\thanks{
This work is part of the scientific program of the
foundation for Fundamental Research on Matter (FOM),
the Dutch Organization for Scientific Research (NWO)
and the TMR program ERB FMRX-CT96-0008.
}
\\
\mbox{}
\\
Division of Physics and Astronomy, Faculty of Sciences,
Vrije Universiteit,\\ De Boelelaan 1081, 1081 HV Amsterdam, The Netherlands
}

\begin{document}


\maketitle

\begin{abstract}
We discuss specific observables that can be measured in deep inelastic
leptoproduction in the case of 1-particle inclusive measurements, namely
azimuthal asymmetries and power-suppressed (higher twist) corrections. 
These quantities contain information on the intrinsic transverse 
momentum of partons, with close connection to the gluon dynamics 
in hadrons.
\end{abstract}

\section{LEPTOPRODUCTION}

\begin{figure}[bth]
\begin{center}
\begin{minipage}{8.5 cm}
\epsfxsize=8.5 cm \epsfbox{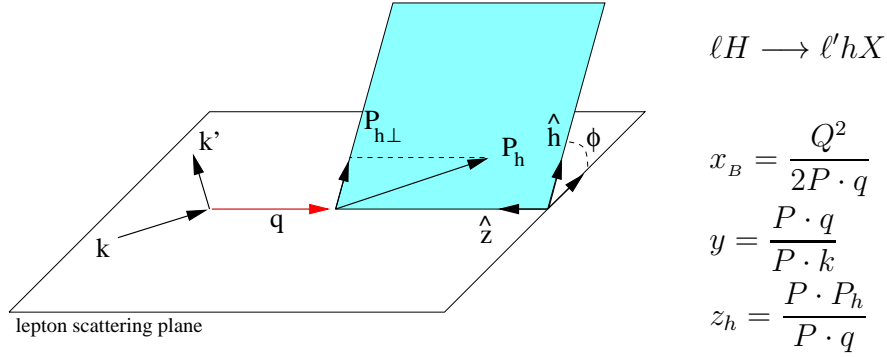}
\end{minipage}
\begin{minipage}{2.5 cm}
\begin{eqnarray*}
&&\ell H \longrightarrow \ell^\prime h X
\end{eqnarray*}
\begin{eqnarray*}
&&\xbj = \frac{Q^2}{2P\cdot q} \\
&&y = \frac{P\cdot q}{P\cdot k}\\
&&z_h = \frac{P\cdot P_h}{P\cdot q}
\end{eqnarray*}
\end{minipage}
\end{center}
\caption{\label{fig1}\em
Kinematics for 1-particle inclusive leptoproduction.}
\end{figure}

The use of polarization in leptoproduction in combination with
azimuthal sensitivity in the final state provides ways to probe
new aspects of hadronic structure. The central
object of interest for 1-particle
inclusive leptoproduction, the hadronic tensor, is given by
\bea
&&2M{\cal W}_{\mu\nu}^{(\ell H)}( q; {P S; P_h S_h} )
=\frac{1}{(2\pi)^4}
\int \frac{d^3 P_X}{(2\pi)^3 2P_X^0}
(2\pi)^4 \delta^4 (q + P - P_X - P_h)
\nonumber
\\
&& \hspace{3.5 cm} \times
\langle {P S} |{J_\mu (0)}|P_X; {P_h S_h} \rangle
\langle P_X; {P_h S_h} |{J_\nu (0)}|{P S} \rangle,
\eea
where $P,\ S$ and $P_h,\ S_h$ are the momenta and spin vectors
of target hadron and produced hadron,
$q$ is the (spacelike) momentum transfer with $-q^2$ = $Q^2$ sufficiently
large. 
The kinematics is illustrated in Fig.~\ref{fig1}, where also the scaling
variables are introduced.
Within the framework of QCD, it is possible to write down a diagrammatic
expansion with the simplest diagrams being given in Fig.~\ref{fig2} for
inclusive and 1-particle inclusive deep inelastic scattering.

\subsection{From hadrons to quarks}

\begin{figure}[b]
\begin{center}
\leavevmode
\epsfxsize=4.5cm \epsfbox{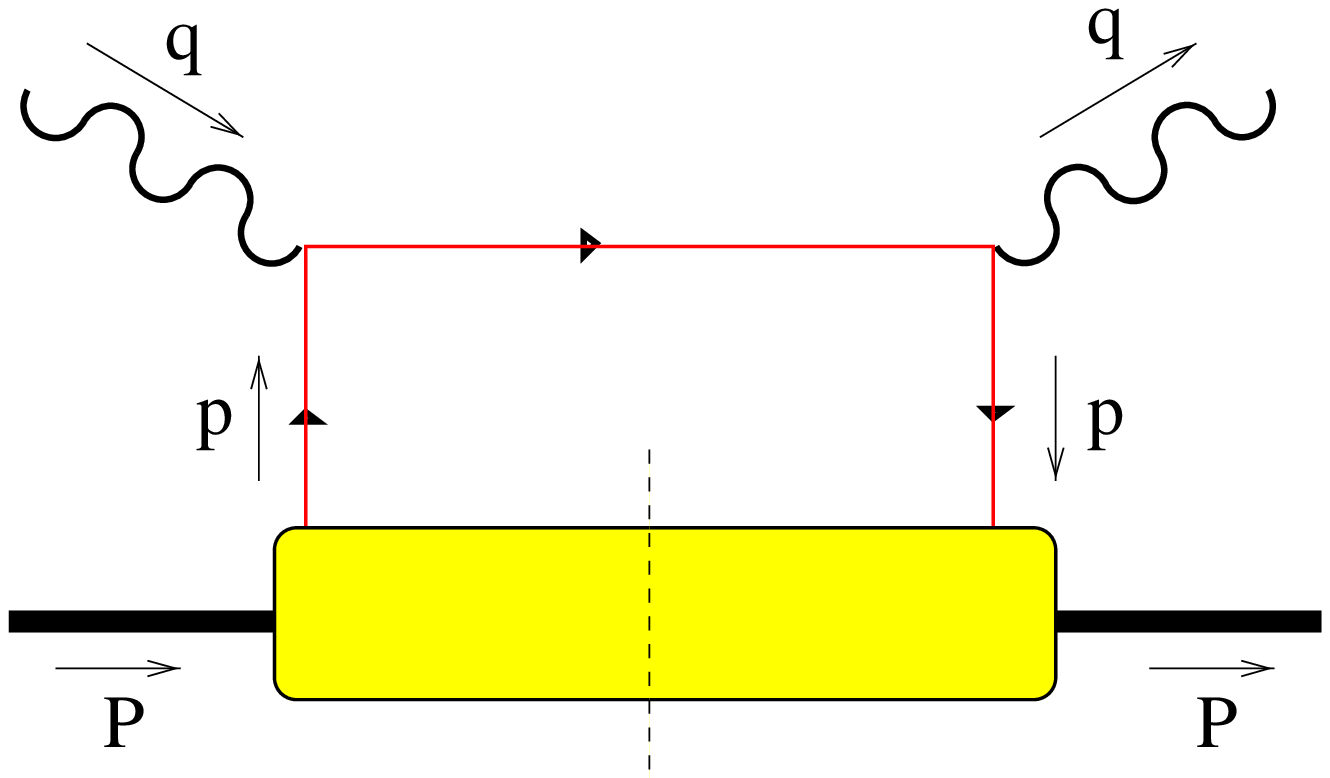}
\hspace{2 cm}
\epsfxsize=4.5cm \epsfbox{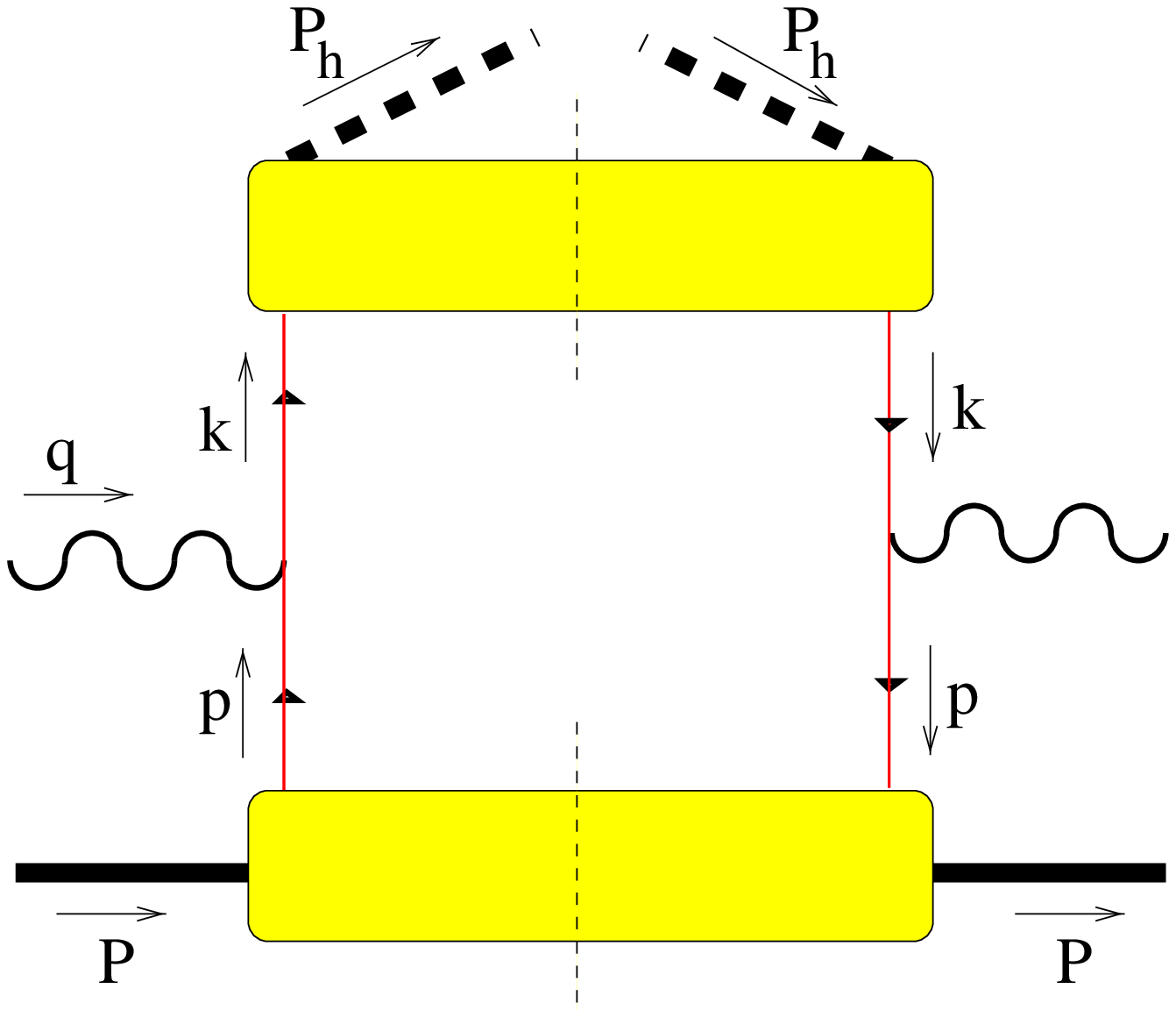}
\end{center}
\caption{\label{fig2}\em
The simplest (parton-level) diagrams representing the squared amplitude
in lepton hadron inclusive scattering (left) and semi-inclusive scattering
(right).}
\end{figure}
In the calculations the relevant structural information for the
hadrons is contained in soft parts (the blobs in Fig.~\ref{fig2})
which represent specific matrix elements of quark fields.
In order to find out which information in the soft parts 
is important in a hard process one needs to realize
that the hard scale $Q$ leads in a natural way to the use of lightlike
vectors $n_+$ and $n_-$ satisfying $n_+^2 = n_-^2 = 0$ and $n_+\cdot n_-$
= 1. For inclusive scattering one parametrizes the momenta
\[
\left.
\begin{array}{l} q^2 = -Q^2 \\
P^2 = M^2\\
2\,P\cdot q = \frac{Q^2}{\xbj} \\
\end{array} \right\}
\longleftrightarrow \left\{
\begin{array}{l}
q =\frac{Q}{\sqrt{2}}\,n_- - \frac{Q}{\sqrt{2}}\,n_+ + q_T
\\ \mbox{} \\
P = \frac{\xbj M^2}{Q\sqrt{2}}\,n_-
+ \frac{Q}{\xbj \sqrt{2}}\,n_+
\end{array}
\right.
\]
The minus component $p^- \equiv p\cdot n_+$
and transverse components are not relevant
in the hard part. The soft part to look at is~\cite{Soper77,Jaffe83}
\be
\Phi_{ij}(x) =
\left. \int \frac{d\xi^-}{4\pi}\ e^{ip\cdot \xi}
\,\langle P,S\vert \overline \psi_j(0) \psi_i(\xi)
\vert P,S\rangle \right|_{\xi^+ = \xi_\st = 0},
\ee
where $x = p^+/P^+$. For the leading order in $1/Q$, it 
is parametrized as~\cite{JJ92}
\be
\Phi(x) =
\frac{1}{4}\,\Biggl\{
f_1(x)\,\nslash_+
+ \lambda\,g_1(x)\, \gamma_5\,\nslash_+
+ h_1(x)\,\frac{\gamma_5\,[\Sslash_\st,\nslash_+]}{2}\Biggr\}
+ {\cal O}\left(\frac{M}{P^+}\right)
\ee
Adding the flavor index $a$, the functions are the 
unpolarized quark distribution $f_1^a$, the chirality distribution
$g_1^a$ and the transverse spin distribution $h_1^a$.
For each of these functions there are many aspects to be discussed,
such as their interpretation (we will come back to this), positivity
and bounds, e.g. $\vert g_1^a(x)\vert \le f_1^a(x)$, symmetry
relations and antiquark distributions, e.g. $\bar f_1(x) = - f_1(-x)$,
sum rules, etc.

\begin{figure}[h]
\leavevmode
\begin{center}
\begin{minipage}{9.0cm}
\epsfxsize=1.8 cm \epsfbox{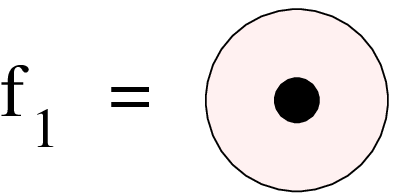}
\hspace{2.5 cm}
\epsfxsize=3.1 cm \epsfbox{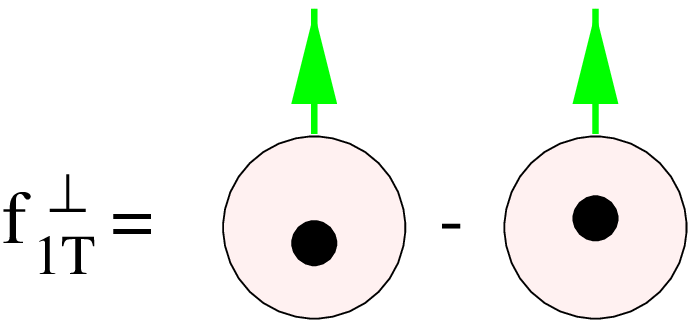}
\\[0.2cm]
\epsfxsize=4.2 cm \epsfbox{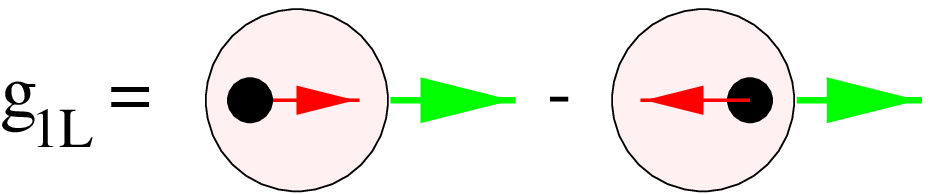}
\hspace{0.5 cm}
\epsfxsize=3.1 cm \epsfbox{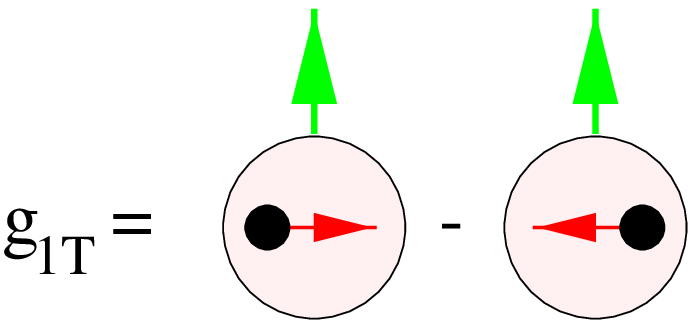}
\\[0.2cm]
\epsfxsize=3.1 cm \epsfbox{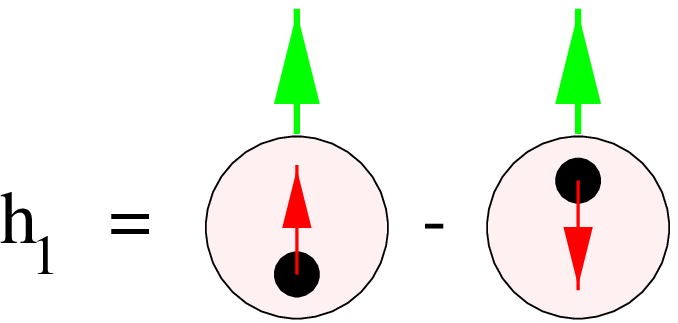}
\hspace{1.3cm}
\epsfxsize=4.2 cm \epsfbox{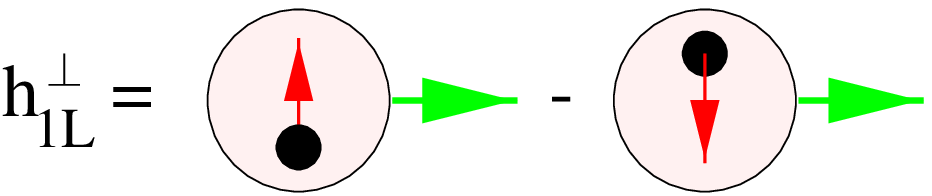}
\\[0.6cm]
\epsfxsize=3.1 cm \epsfbox{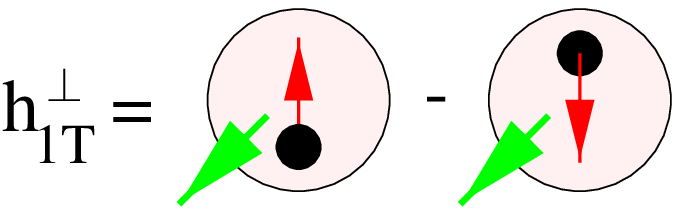}
\hspace{1.3 cm}
\epsfxsize=3.1 cm \epsfbox{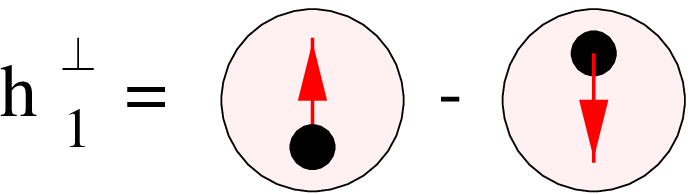}
\end{minipage}
\end{center}
\caption{\em
Interpretation of the functions in the leading
Dirac projections of $\Phi$.
\label{fig3}
}
\end{figure}

For 1-particle inclusive scattering one parametrizes the momenta
\[
\left.
\begin{array}{l} q^2 = -Q^2 \\
P^2 = M^2\\
P_h^2 = M_h^2 \\
2\,P\cdot q = \frac{Q^2}{\xbj} \\
2\,P_h\cdot q = -z_h\,Q^2
\end{array} \right\}
\longleftrightarrow \left\{
\begin{array}{l}
P_h = \frac{z_h\,Q}{\sqrt{2}}\,n_-
+ \frac{M_h^2}{z_h\,Q\sqrt{2}}\,n_+
\\ \mbox{} \\
q =\frac{Q}{\sqrt{2}}\,n_- - \frac{Q}{\sqrt{2}}\,n_+ + q_T
\\ \mbox{} \\
P = \frac{\xbj M^2}{Q\sqrt{2}}\,n_-
+ \frac{Q}{\xbj \sqrt{2}}\,n_+
\end{array}
\right.
\]
The minus component $p^-$ still is not relevant
in the hard part, but the transverse component is. 
The soft part to look at is
\be
\Phi(x,\bm p_T) =
\left. \int \frac{d\xi^-d^2\bm \xi_T}{2\,(2\pi)^3}\ e^{ip\cdot \xi}
\,\langle P,S\vert \overline \psi(0) \psi(\xi)
\vert P,S\rangle \right|_{\xi^+ = 0}.
\ee
For the leading order results, it is parametrized as
\be
\Phi(x,\bm p_\st)  =
\Phi_O(x,\bm p_\st) + \Phi_L(x,\bm p_\st) +\Phi_T(x,\bm p_\st),
\ee
with the parts involving unpolarized
targets (O), longitudinally polarized targets (L) and transversely
polarized targets (T) given by
\bea
\Phi_O(x,\bm p_\st) & = & 
\frac{1}{4} \Biggl\{
f_1(x,\bm p_\st)\,\nslash_+
+ h_1^\perp(x,\bm p_\st)\,\frac{i\,[\pslash_\st,\nslash_+]}{2M}
\Biggr\}
\nonumber \\ 
\Phi_L(x,\bm p_\st) & = & 
\frac{1}{4} \Biggl\{
+ \lambda\,g_{1L}(x,\bm p_\st)\,\gamma_5\,\nslash_+
+ \lambda\,h_{1L}^\perp(x,\bm p_\st)
\Biggr\}
\nonumber \\ 
\Phi_T(x,\bm p_\st) & = & 
\frac{1}{4} \Biggl\{
+ f_{1T}^\perp(x,\bm p_\st)\, \frac{\epsilon_{\mu \nu \rho \sigma}
\gamma^\mu n_+^\nu p_\st^\rho S_\st^\sigma}{M}
+ \frac{\bm p_\st\cdot\bm S_\st}{M}\,g_{1T}(x,\bm p_\st)
\,\gamma_5\,\nslash_+
\nonumber \\ & & \qquad
+ h_{1T}(x,\bm p_\st)\,\frac{\gamma_5\,[\Sslash_\st,\nslash_+]}{2}
+ \frac{\bm p_\st\cdot\bm S_\st}{M}\,h_{1T}^\perp(x,\bm p_\st)\,
\frac{\gamma_5\,[\pslash_\st,\nslash_+]}{2M}
\Biggr\}.
\eea
Again all functions appearing here have a natural interpretation
as densities, now including densities such as the density of
longitudinally polarized quarks in a transversely polarized nucleon
($g_{1T}$) and the density of transversely polarized quarks in
a longitudinally polarized nucleon ($h_{1L}^\perp$). 
These functions vanish from the soft part upon integration
over $p_\st$. Actually we will find that particularly interesting
functions to consider are
\be
g_{1T}^{(1)}(x) = \int d^2p_\st\ \frac{\bm p_\st^2}{2M^2}
\,g_{1T}(x,\bm p_\st),
\ee
and similarly the function $h_{1L}^{\perp (1)}$.
The functions
$h_1^\perp$ and $f_{1T}^\perp$ are T-odd, vanishing if T-reversal
invariance can be applied to the matrix element. For $k_\st$-dependent
correlation functions, matrix elements involving gluonic fields at
infinity (gluonic poles~\cite{bmt}) can for instance prevent 
application of T-reversal invariance. 
The functions describe the possible
appearance of unpolarized quarks in a transversely polarized nucleon
($f_{1T}^\perp$) or transversely polarized quarks in an unpolarized
hadron ($h_1^\perp$) and lead to single-spin asymmetries in various
processes~\cite{Sivers90,Anselmino95}.
The interpretation of the
functions is illustrated in Fig.~\ref{fig3}.

If one proceeds up to order $1/Q$ one also needs to include in
the parametrization of the soft part the parts proportional to
$M/P^+$ and account for gluonic diagrams. For the 
$\bm k_\st$-integrated correlations one has
\begin{eqnarray}
\Phi(x) & = & 
\frac{1}{4}\,\Biggl\{
f_1(x)\,\nslash_+
+ \lambda\,g_1(x)\, \gamma_5\,\nslash_+
+ h_1(x)\,\frac{\gamma_5\,[\Sslash_\st,\nslash_+]}{2}\Biggr\}
\nonumber \\ & &
+ \frac{M}{4P^+}\Biggl\{
e(x) + g_T(x)\,\gamma_5\,\Sslash_\st
+ \lambda\,h_L(x)\,\frac{\gamma_5\,[\nslash_+,\nslash_-]}{2} \Biggr\}
\nonumber \\ & &
+ \frac{M}{4P^+}\Biggl\{
-\lambda\,e_L(x)\,i\gamma_5
- f_T(x)\,\epsilon_\st^{\rho\sigma}\gamma_\rho S_{\st\sigma}
+ h(x)\,\frac{i\,[\nslash_+,\nslash_-]}{2} \Biggr\}.
\end{eqnarray}
Actually gluonic diagrams (what is needed are matrix elements 
containing $\overline \psi(0)\,A_\st^\alpha(\xi)\,\psi(\xi)$) 
do not give
rise to new functions, but they can be related to the above
subleading result using the QCD equations of motion. It is 
important, however, to include gluonic contributions in order to
obtain a gauge invariant result.

{}From Lorentz invariance one obtains, furthermore, some interesting
relations between the subleading functions and the $k_\st$-dependent
leading functions~\cite{BKL84,TM96,BM98}
\bea
&&g_T  = g_1 + \frac{d}{dx}\,g_{1T}^{(1)},
\label{gTrel}
\\
&&h_L = h_1 - \frac{d}{dx}\,h_{1L}^{\perp (1)},
\label{hLrel}
\\
&&f_T =  - \frac{d}{dx}\,f_{1T}^{\perp (1)},
\\
&&h =  - \frac{d}{dx}\,h_{1}^{\perp (1)}.
\eea

\subsection{From quarks to hadrons}

Just as for the distribution functions one can perform an analysis of
the soft part describing the quark fragmentation.
One needs~\cite{CS82}
\be
\Delta_{ij}(z,\bkt) =
\left. \sum_X \int \frac{d\xi^+d^2\bm \xi_\st}{4z\,(2\pi)^3} \,
e^{ik\cdot \xi} \,Tr  \langle 0 \vert \psi_i (\xi) \vert P_h,X\rangle
\langle P_h,X\vert\overline \psi_j(0) \vert 0 \rangle
\right|_{\xi^- = 0}.
\ee
For the production of unpolarized hadrons $h$ one needs in leading order in
$1/Q$ in hard processes~\cite{TM96}
\be
\Delta(z,\bm k_\st) =
\frac{1}{4} \Biggl\{
D_1(z,\bm k^\prime_\st)\,\nslash_-
+ H_1^\perp(z,\bm k^\prime_\st)\,\frac{i\,[\kslash_\st,\nslash_-]}{2M_h}
\Biggr\} + {\cal O}\left(\frac{M_h}{P_h^-}\right).
\ee
The arguments of the fragmentation functions $D_1$ and $H_1^\perp$ are
chosen to be $z$ = $P_h^-/k^-$ and $\bm k^\prime_\st$
= $-z\bkt$. The first
is the (lightcone) momentum fraction of the produced hadron, the second
is the transverse momentum of the produced hadron with respect to the quark.
The fragmentation function $D_1$ is the equivalent of the distribution
function $f_1$. It can be interpreted as the probability of finding a
hadron $h$ in a quark. 
The function $H_1^\perp$, interpretable as the difference in
production probabilities of unpolarized hadrons from a transversely
polarized quark depending on transverse momentum, is allowed 
because of the non-applicability of time reversal invariance. 
This is natural for the fragmentation
functions because of the  appearance of out-states
$\vert P_h, X\rangle$ in the definition of $\Delta$, in contrast
to the plane wave states appearing in $\Phi$.
After $\bkt$-averaging one is left with the functions
$D_1(z)$ and the $\bkt/M$-weighted result $H_1^{\perp (1)}(z)$.
As in the case of distribution functions, the latter function can
be related to a function $H(z)$, appearing at subleading order,
\be
\frac{H(z)}{z} = z^2\,\frac{d}{dz} \left(\frac{H_1^{\perp}}{z}\right).
\label{frag1}
\ee

\section{CROSS SECTIONS FOR LEPTOPRODUCTION}

After the analysis of the soft parts, the next step is to find
out how one obtains the information on the various correlation functions
from experiments, in this particular case in lepton-hadron scattering
via one-photon exchange as discussed before.
To get the leading order result for semi-inclusive scattering it is
sufficient to compute the diagram in Fig.~\ref{fig2} (right)
by using QCD and QED Feynman rules in the hard part and the
matrix elements $\Phi$ and $\Delta$ for the soft parts, parametrized in
terms of distribution and fragmentation functions. The most
well-known results for leptoproduction are:
\newline\newline
\fbox{
\begin{minipage}{15.4cm}
{\bf Cross sections (leading in $1/Q$)}
\bea
&&\frac{d\sigma_{OO}}{d\xbj\,dy\,dz_h}
= \frac{2\pi \alpha^2\,s}{Q^4}\,\sum_{a,\bar a} e_a^2
\left\lgroup 1 + (1-y)^2\right\rgroup \xbj {f^a_1}(\xbj)\,{ D^a_1}(z_h)
\\ && \frac{d\sigma_{LL}}{d\xbj\,dy\,dz_h}
= \frac{2\pi \alpha^2\,s}{Q^4}\,{\lambda_e\,\lambda}
\,\sum_{a,\bar a} e_a^2\  y (2-y)\  \xbj {g^a_1}(\xbj)\,{D^a_1}(z_h)
\eea
\end{minipage}
}
\newline\newline
The indices attached to the cross section refer to polarization
of lepton (O is unpolarized, L is longitudinally polarized) and
hadron (O is unpolarized, L is longitudinally polarized, T is 
transversely polarized). Note that the result is a weighted sum
over quarks and antiquarks involving the charge $e_a$ squared.
Comparing with well-known formal expansions of the cross section in
terms of structure functions one can simply identify these. 
For instance the above result for unpolarized scattering (OO) 
shows that after averaging over azimuthal angles,
only one structure function survives if we work at order $\alpha_s^0$ 
and at leading order in $1/Q$.

It is well-known that in 1-particle inclusive unpolarized leptoproduction
in principle four structures appear, two of them containing azimuthal
dependence of the form $\cos (\phi_h^\ell)$ and $\cos (2\phi_h^\ell)$.
The first one only appears at order $1/Q$~\cite{LM94}, the second one
even at leading order but only in the case of the existence
of nonvanishing T-odd distribution functions. To be specific if we
define weighted cross section such as
\be
\int d^2\bm q_{T}\,\frac{Q_{T}^2}{MM_h} \,\cos(2\phi_h^\ell)
\,\frac{d\sigma_{{OO}}}{d\xbj\,dy\,dz_h\,d^2\bm q_{T}}
\equiv
\left< \frac{Q_{T}^2}{MM_h} \,\cos(2\phi_h^\ell)\right>_{OO}
\ee
we obtain the following asymmetry.
\vspace{0.1 cm}\newline\fbox{
\begin{minipage}{15.4cm}
{\bf Azimuthal asymmetries for unpolarized targets (leading twist)}
\be
\left<
\frac{Q_{T}^2}{MM_h} \,\cos(2\phi_h^\ell)\right>_{OO}
= \frac{16\pi \alpha^2\,s}{Q^4}
\,(1-y)\,\sum_{a,\bar a} e_a^2
\,\xbj\,{h_{1}^{\perp(1)a}}(\xbj) H_1^{\perp (1)a}.
\ee
\end{minipage}
}
\newline\newline
An interesting asymmetry involving the same fragmentation part is
a $\sin(\phi_h^\ell)$ single spin asymmetry, requiring only a polarized
lepton but no polarization for the target~\cite{LM94}.
\newline\newline
\fbox{
\begin{minipage}{15.4cm}
{\bf Single spin asymmetry for unpolarized targets (higher twist)}
\bea
&&\left< \frac{Q_{T}} {M} \,\sin(\phi_h^\ell) \right>_{LO}
= \frac{4\pi \alpha^2\,s}{Q^4}\,{\lambda_e}
\,y\sqrt{1-y} 
\,\frac{2M}{Q}\,\xbj^2 {\tilde e^a}(\xbj)\,{H_1^{\perp (1)a}}(z_h)
\label{as1}
\\ &&
\mbox{note:}
\ {\tilde e^a}(x) = e^a(x) - \frac{m_a}{M}\,\frac{f_1^a(x)}{x}.
\nonumber
\eea
\end{minipage}
}
\newline\newline
This cross section involves, besides the 
time-reversal odd fragmentation function $H_1^\perp$, 
the distribution function $e$.
The tilde function that appear in the cross sections is in fact
the socalled interaction dependent part of the twist three
functions. It would vanish in any naive parton model calculation in
which cross sections are obtained by folding electron-parton cross
sections with parton densities. Considering the relation for $\tilde e$
one can state it as $x\,e(x)$ = $(m/M)\,f_1(x)$ in the absence of
quark-quark-gluon correlations. The inclusion of the latter also
requires diagrams dressed with gluons.

For polarized targets, several azimuthal asymmetries arise already
at leading order. For example the following possibilities were
investigated in Refs~\cite{KM96,Collins93,Kotzinian95,TM95b}.
\newline \newline
\fbox{
\begin{minipage}{15.4 cm}
{\bf Azimuthal asymmetries for polarized targets (leading twist)}
\ba
&&
\left< \frac{Q_\st}
{M} \,\cos(\phi_h^\ell-\phi_S^\ell)\right>_{LT}
= \frac{2\pi \alpha^2\,s}{Q^4}\,{\lambda_e\,\vert \bSt \vert}
\,y(2-y)\sum_{a,\bar a} e_a^2
\,\xbj\,{g_{1T}^{(1)a}}(\xbj) {D^a_1}(z_h),
\\
&&
\left< \frac{Q_\st^2}{MM_h}
\,\sin(2\phi_h^\ell)\right>_{OL}
= -\frac{4\pi \alpha^2\,s}{Q^4}\,{\lambda}
\,(1-y)\sum_{a,\bar a} e_a^2
\,\xbj\,{h_{1L}^{\perp(1)a}}(\xbj) {H_1^{\perp(1)a}}(z_h),
\label{as2}
\\
&&
\left< \frac{Q_\st}{M_h}
\,\sin(\phi_h^\ell+\phi_S^\ell)\right>_{OT}
= \frac{4\pi \alpha^2\,s}{Q^4}\,{\vert \bSt \vert}
\,(1-y)\sum_{a,\bar a} e_a^2
\,\xbj\,{h_1^a}(\xbj) {H_1^{\perp(1)a}}(z_h).
\label{finalstate}
\ea
\end{minipage}
}
\newline \newline
The latter two are single spin asymmetries involving the fragmentation
function $H_1^{\perp (1)}$. The last one was the asymmetry proposed by
Collins~\cite{Collins93} as a way to access the transverse spin 
distribution function $h_1$ in pion production. 
Note, however, that in using the
azimuthal dependence one needs to be very careful. For instance, besides
the $<\sin (\phi_h^\ell + \phi_S^\ell)>_{OT}$, one also finds at leading 
order a $<\sin (3\phi_h^\ell - \phi_S^\ell)>_{OT}$ asymmetry which is 
proportional to $h_{1T}^{\perp (2)}\,H_1^{\perp (1)}$~\cite{TM95b}.

\begin{figure}[t]
\begin{minipage}{7.5cm}
\epsfig{file=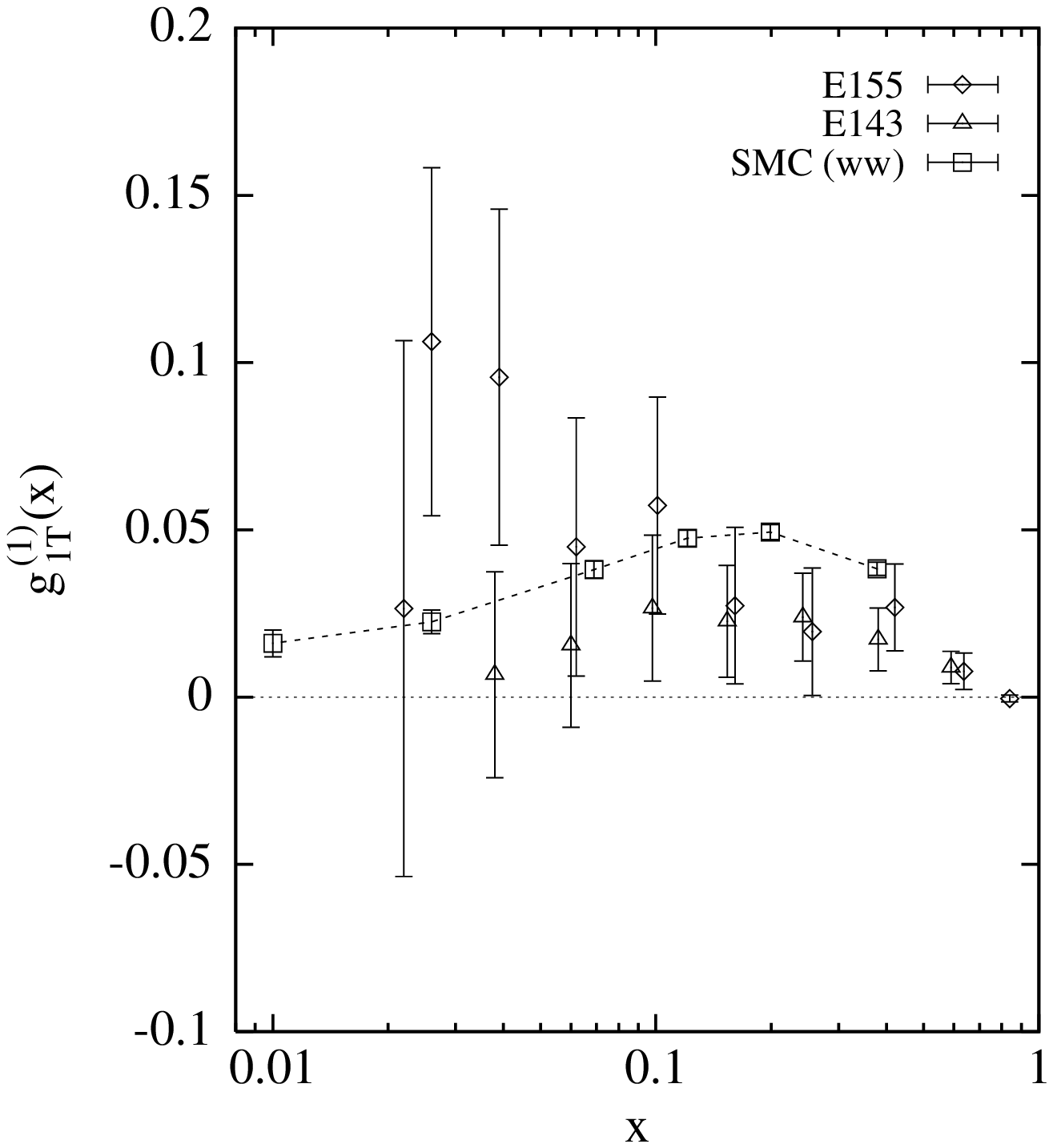,angle=0,width=7cm}
\vspace{-1.0cm}
\caption{\label{g1Tupdat} An estimate for the function $g_{1T}^{(1)}(x)$
using $g_2$-data and the Wandzura-Wilczek approximation obtained from
the SMC $g_1$-data.}
\end{minipage}
\hspace{0.5cm}
\begin{minipage}{7.5cm}
\epsfig{file=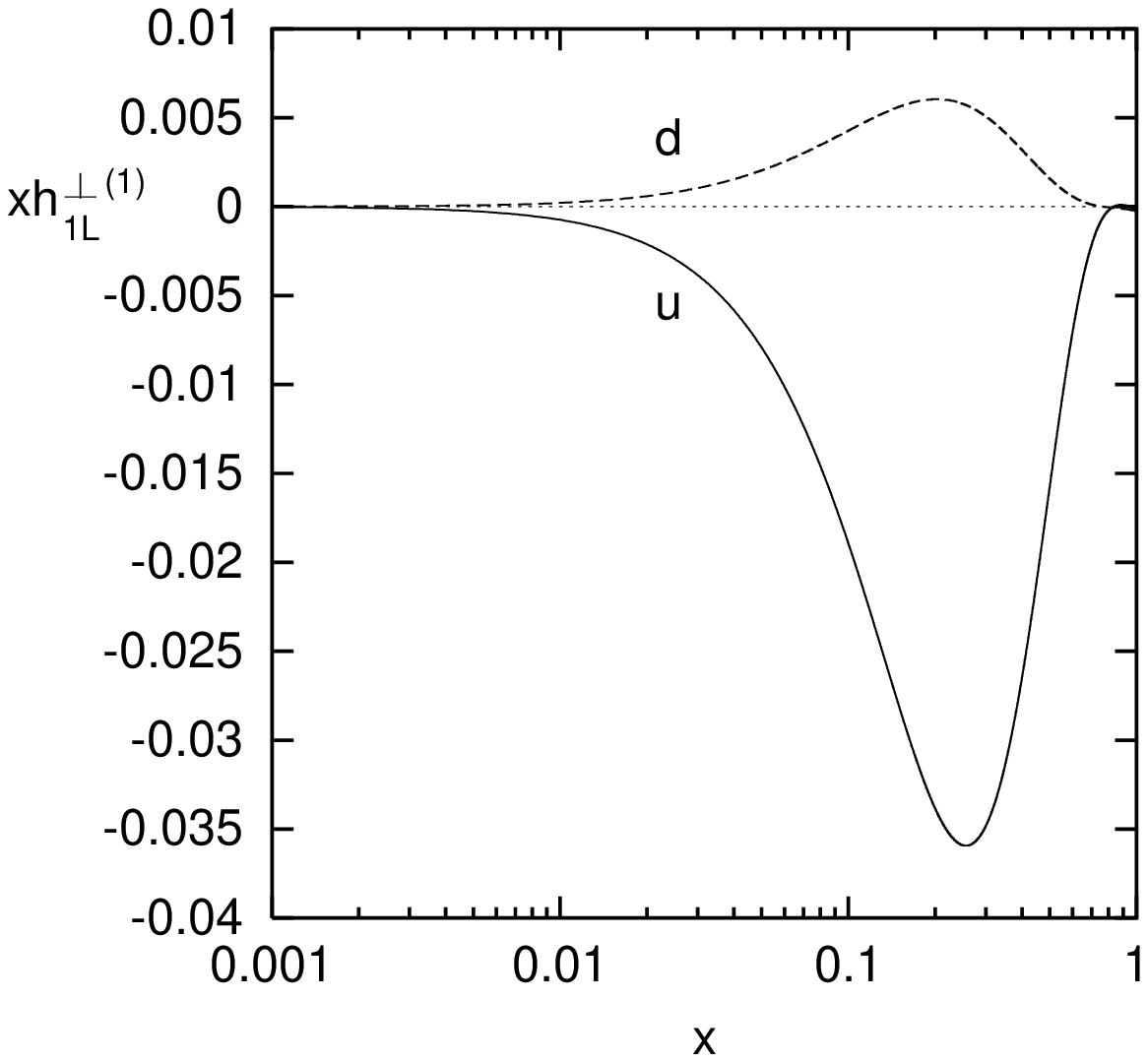,angle=-90,width=7cm}
\vspace{0.1cm}
\caption{\label{h1Lperp1} An estimate for the function 
$h_{1L}^{\perp(1)}(x)$ obtained from $h_1$
using a 'Wandzura-Wilczek'-like approximation.}
\end{minipage}
\vspace{-0.6cm}
\end{figure}

Notice that an estimate of the function $g_{1T}^{(1)}$ can be obtained
from inclusive results for $g_2$ using the relation~\ref{gTrel} for
$g_T = g_1 + g_2$. Actually this relation is exact
and the data from the SLAC E143 experiment were used for an estimate 
in Ref.~\cite{KM96}. An update using the preliminary $g_2$ data from the 
E155 experiment is shown in Fig.~\ref{g1Tupdat}. More about the final
analysis of these data will be reported in Ref.~\cite{Bosted}.
A well-known approximation for $g_2$ is the Wandzura-Wilczek 
result~\cite{WW}. An alternative derivation of this relation is
obtained by combining 
the separation of the twist three function $g_T$ into a twist-two
part and an interaction dependent part $\tilde g_T$, omitting quark
mass terms given by
\be
g_T(x) = \frac{g_{1T}^{(1)}(x)}{x} + \tilde g_T(x) .
\ee
By eliminating $g_{1T}^{(1)}$ one then derives the relation
\be
g_T(x) = \int_x^1 dy\,\frac{g_1(y)}{y}
+ \underbrace{\left(\tilde g_T(x) -
\int_x^1 dy\,\frac{\tilde g_T(y)}{y}\right)}_{\bar g_T(x)}.
\ee
The Wandzura-Wilczek part is obtained by putting $\bar g_T(x) = 0$. 
Within this
approximation one can find an estimate for $g_{1T}^{(1)}$ from the
$g_1$-data, the result of which using the SMC data~\cite{SMC} is 
shown in Fig.~\ref{g1Tupdat}.

In the same experiment in which one would measure the single spin 
asymmetry $<\sin (\phi_h^\ell + \phi_S^\ell)>_{OT}$, one  can also 
measure the possible existence of a
$<\sin (\phi_h^\ell - \phi_S^\ell)>_{OT}$ asymmetry. Such an asymmetry
would arise from a T-odd distribution function; to be precise
\vspace{0.1 cm}\newline\fbox{
\begin{minipage}{15.4cm}
\be
\left<\frac{Q_T}{M_h}\,\sin(\phi^\ell_h-\phi^\ell_S)\right>_{OTO}
= \frac{2\pi \alpha^2\,s}{Q^4}\,\vert \bm S_T \vert
\,\left( 1-y-\frac{1}{2}\,y^2\right) 
\sum_{a,\bar a} e_a^2
\,\xbj\,f_{1T}^{\perp (1)a}(\xbj) D_1^{a}(z_h).
\label{initialstate}
\ee
\end{minipage}
}
\newline\newline
By comparing the results for~\ref{finalstate} and \ref{initialstate},
leptoproduction could resolve an ambiguity in the explanation of the
single spin (left-right)
asymmetry observed in $p^{\uparrow}p \rightarrow \pi X$, which  can 
be attributed
to a T-odd effect in the initial state 
(Sivers effect, $f_{1T}^{\perp (1)}$~\cite{Sivers90,Anselmino95}) 
or a T-odd effect in the final state 
(Collins effect, $H_1^{\perp (1)}$~\cite{Anselmino99}). 
Estimates for the leptoproduction asymmetries for both cases have
been presented in Ref.~\cite{BM99} and are found to have quite
characteristic behavior as a function of $\xbj$ and $z_h$.

\begin{figure}[t]
\begin{minipage}{7.5cm}
\epsfig{file=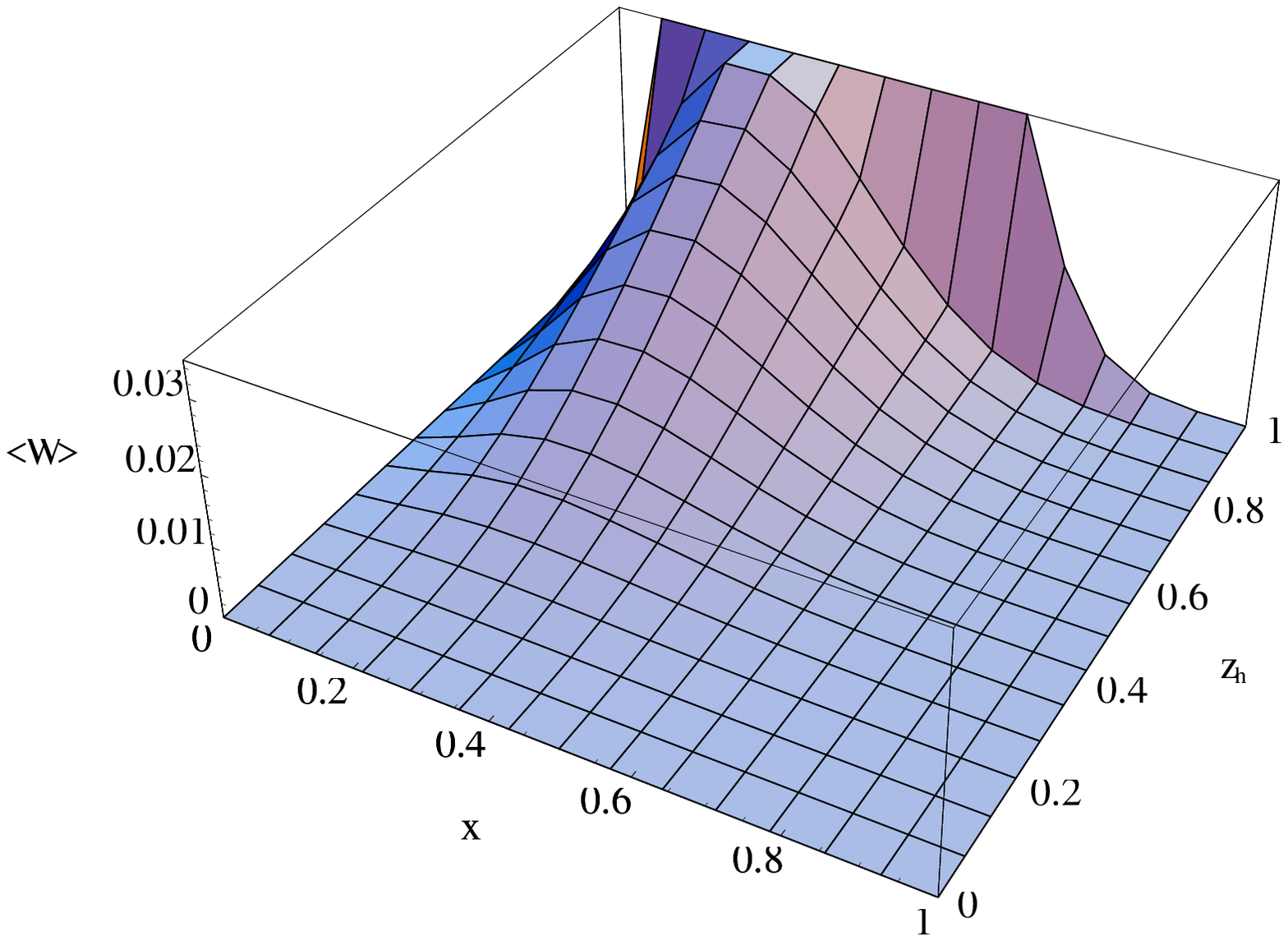,angle=0,width=7cm}
\caption{\label{ws1} An estimate for the asymmetry
$< \sin (\phi_h^\ell)>_{OL}$ (see text). We only plotted the
weighted product of distribution and fragmentation functions
omitting the $y$-dependence.}
\end{minipage}
\hspace{0.5cm}
\begin{minipage}{7.5cm}
\epsfig{file=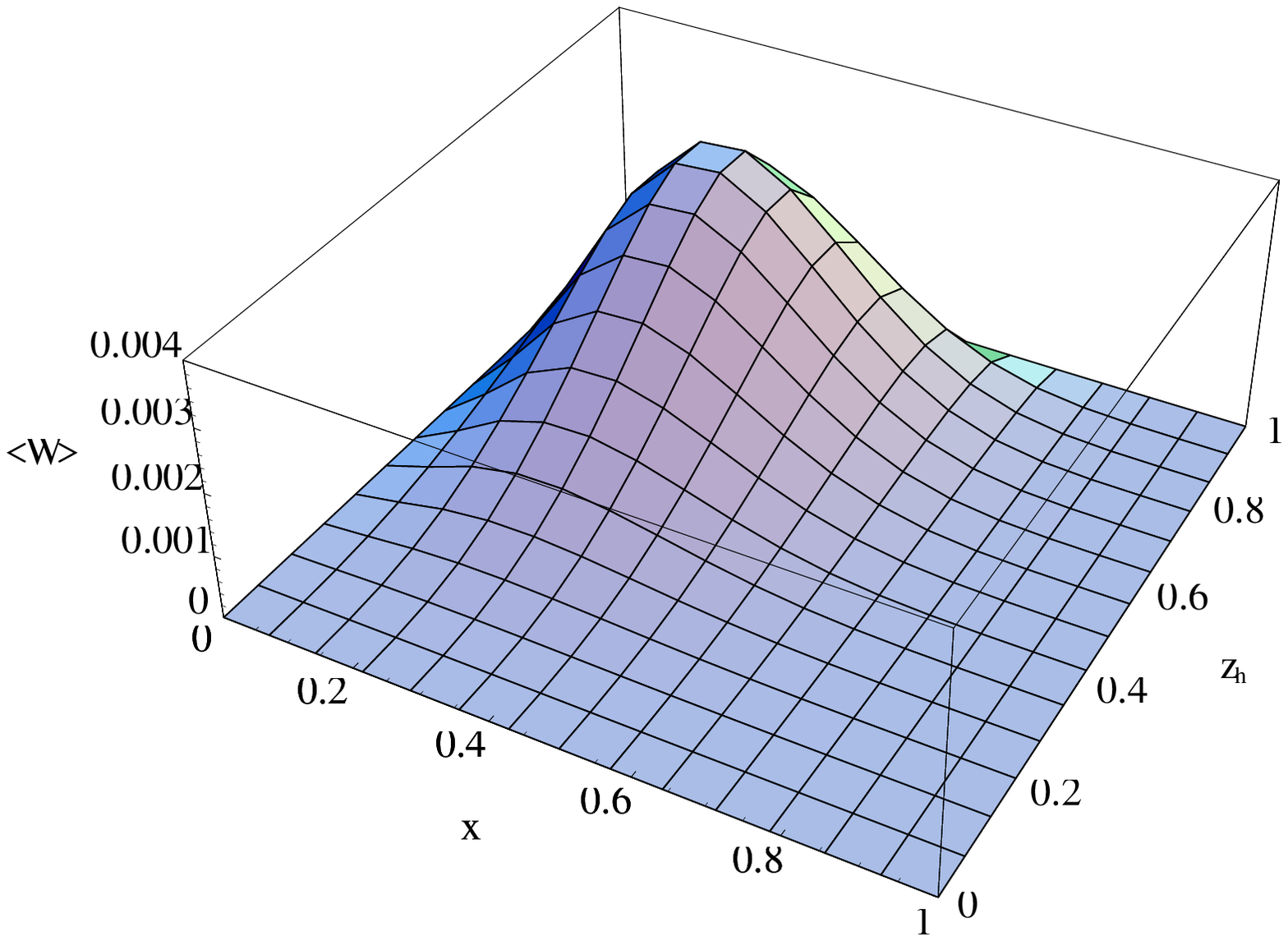,angle=0,width=7cm}
\caption{\label{ws2} An estimate for the asymmetry
$< \sin (2\phi_h^\ell)>_{OL}$ (see text). We only plotted the
weighted product of distribution and fragmentation functions
omitting the $y$-dependence.}
\end{minipage}
\end{figure}

As a final example, we want to concentrate on single spin
asymmetries $< \sin (\phi_h^\ell)>_{LO}$,
$< \sin (\phi_h^\ell)>_{OL}$ and
$< \sin (2\phi_h^\ell)>_{OL}$, for which preliminary results
have been presented~\cite{Oganessyan} by the HERMES collaboration.
The results for two of these have been mentioned already 
(Eqs~\ref{as1} and \ref{as2}), the other one is actually a
higher twist result~\cite{TM96},
\vspace{0.1 cm}\newline\fbox{
\begin{minipage}{15.4cm}
\bea
&&\left<\frac{Q_T}{M_h}\,\sin(\phi^\ell_h)\right>_{OL}
= \frac{4\pi \alpha^2\,s}{Q^4}\,\lambda
\,(2-y)\sqrt{1-y}\, \sum_{a,\bar a} e_a^2
\,\Biggl\{
\xbj\,h_{1L}^{\perp (1)a}(\xbj) \frac{\tilde H^{a}(z_h)}{z_h}
\nonumber \\ && \mbox{} \hspace{5.5cm}
+ \xbj \left(2\,h_{1L}^{\perp (1)a}(\xbj)-\xbj\,\tilde h_L^a(\xbj)\right)
H_1^{\perp (1) a}(z_h)\Biggr\} .
\eea
\end{minipage}
}
\newline\newline
In the same way as discussed for $g_T$ and $g_{1T}^{(1)}$, we also have
relations between the $h$-functions. One can combine the decomposition
of the function $h_L$, appearing at subleading order in leptoproduction,
into a twist-two part and an interaction dependent part (omitting
quark mass dependent terms),
\be
h_L(x) = -2\,\frac{h_{1L}^{\perp (1)}(x)}{x} + \tilde h_L(x),
\ee
with Eq.~\ref{hLrel} to eliminate $h_{1L}^{\perp (1)}$ and find
\be
h_L(x)  = 2x\int_x^1 dy\,\frac{h_1(y)}{y^2}
+ \underbrace{\left(\tilde h_L(x) -
2x\int_x^1 dy\,\frac{\tilde h_L(y)}{y^2}\right)}_{\bar h_L}.
\ee
The approximation $\bar h_L = 0$ e.g. can be used to obtain an idea
of the magnitude of $h_L$ and then with the exact Eq.~\ref{hLrel}
an estimate for $h_{1L}^{\perp (1)}$. One needs as input some reasonable 
guess for $h_1$ (e.g.  in our case we took $h_1$ = $f_1$ up to some 
spin factors) to get the estimate in Fig.~\ref{h1Lperp1}.
We find that the functions $g_{1T}^{(1)}$ and $h_{1L}^{\perp (1)}$ are
of the same order of magnitude, and about an order of magnitude
smaller than the functions $f_1$ and $g_1$. With the presented
estimates and the estimate for $H_1^{\perp(1)}$ from 
Ref.~\cite{Anselmino99} which not only determines the functions $H$ 
(Eq.~\ref{frag1}) but also $\tilde H$,
\be
H(z) =  -2z\,H_1^{\perp (1)}(z) + \tilde H(z),
\ee
we find for the single spin asymmetries the results in
Figs~\ref{ws1} and \ref{ws2}. Note that the 
$< \sin (\phi_h^\ell)>_{OL}$ asymmetry is larger than the
$< \sin (2\phi_h^\ell)>_{OL}$ asymmetry. This shows e.g. that the
absence in the HERMES results of a clear signal for the second
asymmetry does not allow conclusions on the magnitude of 
$h_{1L}^{\perp (1)}$.

\section{CONCLUDING REMARKS}

In the previous section some results for
1-particle inclusive lepton-hadron scattering have been presented. 
Several other effects are important in these cross sections, such as
target fragmentation, the inclusion of gluons in the calculation to
obtain color-gauge invariant definitions of the correlation functions and an
electromagnetically gauge invariant result at order $1/Q$, and finally
QCD corrections which can be moved back and forth between hard
and soft parts, leading to the scale dependence of the soft parts and
the DGLAP equations.

In this contribution we have tried to indicate why semi-inclusive,
in particular 1-particle inclusive
lepton-hadron scattering, can be important. 
The goal is the study of the quark
and gluon structure of hadrons, emphasizing the
dependence on transverse momenta of quarks. The reason why this prospect is
promising is the existence of a field theoretical framework that allows
a clean study involving well-defined hadronic matrix elements. It does 
require, however, also a dedicated experimental effort using polarized
beams, targets and detection of final state hadrons.

\end{document}